\documentclass{emulateapj}

\newcommand{\Ch}{{\em Chandra }}

\def\gappeq{\mathrel{ \rlap{\raise.5ex\hbox{$>$}}
                      {\lower.5ex\hbox{$\sim$}}  } }
\def\lappeq{\mathrel{ \rlap{\raise.5ex\hbox{$<$}}
                      {\lower.5ex\hbox{$\sim$}}  } }

\shorttitle{THE JET/COMPANION-GALAXY INTERACTION IN 3C\,321}
\shortauthors{EVANS ET AL.}

\begin{document}

\title{A Radio Through X-ray Study of the Jet/Companion-Galaxy Interaction in 3C\,321}
\author{Daniel~A.~Evans\altaffilmark{1,2}, Wen-Fai~Fong\altaffilmark{1,3}, Martin~J.~Hardcastle\altaffilmark{4}, Ralph~P.~Kraft\altaffilmark{2}, Julia~C.~Lee\altaffilmark{1,2}, Diana~M.~Worrall\altaffilmark{5}, Mark~Birkinshaw\altaffilmark{5}, Judith~H.~Croston\altaffilmark{4}, Tom~W.~B.~Muxlow\altaffilmark{6}}
\altaffiltext{1}{Harvard University, Department of Astronomy, 60 Garden Street, Cambridge, MA 02138}
\altaffiltext{2}{Harvard-Smithsonian Center for Astrophysics, 60 Garden Street, Cambridge, MA 02138}
\altaffiltext{3}{Massachusetts Institute of Technology, Department of Physics, 77 Massachusetts Avenue, Cambridge MA 02139}
\altaffiltext{4}{School of Physics, Astronomy \& Mathematics, University of Hertfordshire, College Lane, Hatfield AL10 9AB, UK}
\altaffiltext{5}{University of Bristol, Department of Physics, Tyndall Avenue, Bristol BS8 1TL, UK}
\altaffiltext{6}{MERLIN/VLBI National Facility, Jodrell Bank Observatory, University of Manchester, Macclesfield, Cheshire SK11 9DL, UK}

\begin{abstract}

We present a multiwavelength study of the nucleus, environment, jets,
and hotspots of the nearby FRII radio galaxy 3C\,321, using new and
archival data from MERLIN, the VLA, {\it Spitzer}, {\it HST}, and {\it
Chandra}. An initially collimated radio jet extends northwest
from the nucleus of its host galaxy and produces a compact knot of
radio emission adjacent (in projection) to a companion galaxy, after which it
dramatically flares and bends, extending out in a diffuse structure
35~kpc northwest of the nucleus. We argue that the simplest
explanation for the unusual morphology of the jet is that it is
undergoing an interaction with the companion galaxy. Given that the northwest hotspot that lies
$\gappeq$250 kpc from the core shows X-ray emission, which likely
indicates {\it in situ} high-energy particle acceleration, we argue
that the jet-companion interaction is not a steady-state situation.
Instead, we suggest that the jet has been disrupted on a timescale
less than the light travel time to the end of the lobe, $\sim 10^6$
years, and that the jet flow to this hotspot will only be disrupted
for as long as the jet-companion interaction takes place. The host galaxy of
3C\,321 and the companion galaxy are in the process of merging, and
each hosts a luminous AGN. As this is an unusual situation, we
investigate the hypothesis that the interacting jet has driven
material on to the companion galaxy, triggering its AGN. Finally, we
present detailed radio and X-ray observations of both hotspots, which
show that there are multiple emission sites, with spatial offsets
between the radio and X-ray emission.

\end{abstract}


\section{INTRODUCTION}
\label{intro}

Radio galaxies consist of twin jets of particles that are ejected from
a compact active galactic nucleus (AGN), feeding
into large-scale `plumes' or `lobes'. There are two distinct
morphological classes of radio galaxies, low-power (Fanaroff-Riley
type I, hereafter FRI) sources and high-power (FRII) sources
(\citealt{fr74}). FRI sources exhibit `edge-darkened' large-scale
radio structure, with an initially bright radio jet that appears to flare into large plumes on kiloparsec scales. On the other hand, FRII sources appear `edge-brightened', and in these cases relativistic jets propagate out to large distances (often $>$100 kpc) from the core before terminating in bright hotspots and accompanying radio lobes. Observationally, the Fanaroff-Riley divide occurs at a 178-MHz radio power of $\sim2\times10^{25}$ W~Hz$^{-1}$~sr$^{-1}$.

The Fanaroff-Riley classification of an object is influenced by the physics of the interaction between the kiloparsec-scale jets and the external medium through which they propagate (e.g., \citealt{bic95,hub06}). In FRI sources there is evidence (see, e.g., \citealt{lb02}) that the jets are in direct contact with the external medium, and entrainment of material causes the jets to decelerate from relativistic to sub-relativistic speeds over scales of 1--10 kpc. Detailed radio and optical images of FRI jets often show a `knotty' structure with many compact features (\citealt{owen89}) that may be hydrodynamical in nature and related to internal shocks. There is also evidence that the FRI jet in Minkowski's Object has been partially disrupted by a collision with an external cloud of gas (\citealt{cro06}). In the X-ray, \Ch routinely detects X-ray synchrotron emission from FRI jets (e.g., \citealt{wor01}), requiring {\it in situ} particle acceleration.

Jet--environment interactions in the more powerful FRII-type radio galaxies play less of a role in gradual jet deceleration than in FRI-type sources. The jets in FRII sources remain relativistic, often out to scales of hundreds of kpc or more, until they decelerate abruptly (i.e., on scales that are much less than the length scale of the jet), giving rise to bright hotspots. Many FRII radio galaxies and quasars show knotty structure in their jets (e.g., \citealt{vbd93,bri94,best97}). However, there is little evidence to suggest that the knots are associated with major alterations to the propagation of the jets: the jets normally remain well-collimated and highly relativistic all the way to the compact hotspots. Even when jet-environment interactions are thought to be present from other observations, there is seldom any evidence that these disrupt the jet. For example, the sources 3C\,285 (\citealt{vbd93}) and 3C\,34 (\citealt{best97}) both show a series of compact features along their collimated radio jets, while optical observations show regions of supposed jet-induced star formation. However, the large spatial offsets ($>$15~kpc) between the optical star-forming regions and the radio knots are difficult to reconcile in jet--environment models.

3C\,321 ($z$=0.0961, $D_{\rm L}$=441 Mpc) is a prominent example of a FRII radio source that not only shows evidence (e.g., \citealt{baum88}) for a radio knot, but {\it also} has a one-sided, diffuse (i.e., FRI-like) jet. On small scales, a radio knot lies $\sim$3.5$''$ ($\sim$6.2~kpc) northwest of an unresolved core, and a seemingly bent and diffuse jet extends between $\sim$13~kpc and $\sim$30~kpc. On larger scales, two compact hotspots are detected $\sim$145$''$ ($\sim$260~kpc) from the nucleus at both radio and X-ray wavelengths. \cite{har04} demonstrated that the X-ray emission from the hotspots is most likely synchrotron in nature, which implies on-going particle acceleration in the form of a well-collimated energy input.

Optically, observations of the nuclear regions of 3C\,321 with {\it HST} (e.g.,~\citealt{hur99,mar99,dekoff00}) show two components, with the brighter, southeast member crossed by a kpc-scale dust lane, which itself shows a series of filaments. The absolute V-band magnitude of the southeast and northwest components is -22.3 and -21.2, respectively (\citealt{roche00}). \cite{baum88} associate the compact radio nucleus with the southeast component. The northwest component is a companion galaxy, as is demonstrated by the common stellar envelope of the two sources (e.g., \citealt{hur99}) and the consistency to within 200 km~s$^{-1}$ of their redshifts (\citealt{rob00}). Further, the optical emission-line spectra of the main galaxy and its companion (e.g., \citealt{fil87,rob00}) indicate that {\it both} may host AGN. If confirmed in the X-ray, 3C\,321 would be a rare example of two merging galaxies, each of which hosts an active nucleus. To date, the best example in the literature of this situation is in the $z$=0.0244 system NGC 6240 (\citealt{kom03}) -- clear evidence for the presence of AGN here comes from the detection of Fe K$\alpha$ lines in both galaxies. There is a handful of other examples of possible binary AGN, but in these cases the second galaxy tends to be significantly less luminous than its partner so that AGN emission cannot be unambiguously distinguished from starburst activity (e.g., \citealt{bal04,jim07}).

The compact knot and bright one-sided jet in 3C\,321 make it an excellent target for a detailed study of the role of environmental interactions in FRII jets. We argue in this paper, based on new and archival data from MERLIN, the VLA, {\it Spitzer}, {\it HST}, and {\it Chandra}, that the jet has undergone an interaction with the companion galaxy that lies immediately to the northwest of the host galaxy of 3C\,321. This paper is organized as follows. Section 2 contains a description of the data and a summary of their analysis. In Section 3, we present the results of our observations of the nucleus, environment, jets, and hotspots of 3C\,321. In Section 4, we present evidence that the companion galaxy has temporarily disrupted the propagation of the jet, and in Section 5 we describe the emission mechanisms involved in generating the complex emission in the north and south hotspots in the source. We end with our conclusions in Section 6.

\section{OBSERVATIONS AND ANALYSIS}

\subsection{Radio}

We used archival VLA observations (Table \ref{vlaobs}) to make images of the source at 1.5 and 4.8 GHz at various resolutions. Our maps are of comparable quality to those produced by \cite{baum88}. The data were reduced in {\sc aips} in the standard manner. In the case of the 4.8-GHz data, we cross-calibrated and merged datasets from the three VLA configurations to produce a single dataset that sampled both the small- and large-scale structure of the radio galaxy.

We observed 3C\,321 with MERLIN on 2005 June 20 in line mode at two frequencies, 1408 and 1658 MHz, over a period of $\sim 14$ hours. The MERLIN data were processed in the standard way and then imported into {\sc aips} for self-calibration. After self-calibration and merging of the two MERLIN frequencies the dataset was then concatenated with the A-configuration VLA data described in Table \ref{vlaobs} to provide short-baseline sensitivity. The resulting dataset has an effective frequency of $\sim 1.5$ GHz.

\subsection{Spitzer}

3C\,321 was observed with the {\it Spitzer} IRAC instrument on 2005 March 27 (P.I. M. Birkinshaw), with an AOR duration of 1,072~s. The data used were the Post-Basic Calibrated Data (PBCD) files, which include a flux-calibrated mosaic (`MAIC') file for each channel that is suitable for basic scientific analysis.

\subsection{HST}

We used three archival {\it HST} WFPC2 observations of 3C\,321. The first was taken on 1995 April 29 with the F702W filter (280~s exposure). The second was taken on 1995 August 12 with the FR533NR linear ramp filter tuned to 5487.7\AA\ (corresponding to [O~{\sc iii}] in the rest frame of 3C\,321), with a 600~s exposure. Finally, we used an STIS NUV-MAMA observation taken on 2000 June 05 (1,440~s exposure).

\subsection{Chandra}

3C\,321 was observed with the \Ch ACIS CCD camera on 2002 April 30 (OBSID 3138).  The observation was made in FAINT mode, with the source placed near the standard aimpoint of the S3 chip.  We reprocessed the data using {\sc CIAO} v3.3.0.1 with the CALDB 3.2.1 calibration database to create a new level-2 events file filtered for the ASCA grades 0, 2, 3, 4, and 6 and with the 0.5$''$ pixel randomization removed. To check for periods of high background, we extracted lightcurves for the entire S3 chip, excluding point sources. Inspection of the lightcurves showed minor flaring, and subsequent GTI-filtering reduced the exposure time from 47,135 s to 46,932 s.

\subsection{Astrometry and the positions of the nuclei}

We compared the relative alignment of the datasets, and found that
small astrometric corrections of order 1$''$ were needed. We used the
unresolved radio core as a reference point, and compared it to a
hard-band (4--7 keV) \Ch image. We found the radio nucleus is
well-aligned with the southwest component of X-ray emission. The 90\%
uncertainty of the \Ch X-ray absolute
position\footnote{http://cxc.harvard.edu/cal/ASPECT/celmon/} has a
radius of 0.6$''$, so this gives us confidence that the radio and
X-ray components are associated with one another. The {\it HST}
astrometric uncertainty tends to be somewhat greater, but fortunately
the dust lane that crosses the southwest galaxy is also detected in
absorption in the \Ch image, which allows us to align the datasets.

\section{RESULTS}

All results presented in this paper use a cosmology in which
$\Omega_{\rm m, 0}$ = 0.3, $\Omega_{\rm \Lambda, 0}$ = 0.7, and H$_0$
= 70 km s$^{-1}$ Mpc$^{-1}$. At the redshift of 3C\,321 ($z$ = 0.0961),
1 arcsec corresponds to 1.78 kpc. All spectral fits include absorption
through our Galaxy using $N_{\rm H, Gal}$ = 4.14$\times$10\(^{20}\)
cm$^{-2}$ (\citealt{dic90}). Errors quoted in this paper are 90 per
cent confidence for one parameter of interest (i.e., $\chi^2_{\rm
  min}$ + 2.7), unless otherwise stated. The radio spectral index, $\alpha$, is defined by the relation $S_\nu=\nu^{-\alpha}$, where $S_\nu$ is the flux density and $\nu$ is the frequency.

\subsection{The Small-Scale Jet}
\label{ssjet}

\begin{figure}
\begin{center}
\includegraphics[angle=0,width=8.5cm]{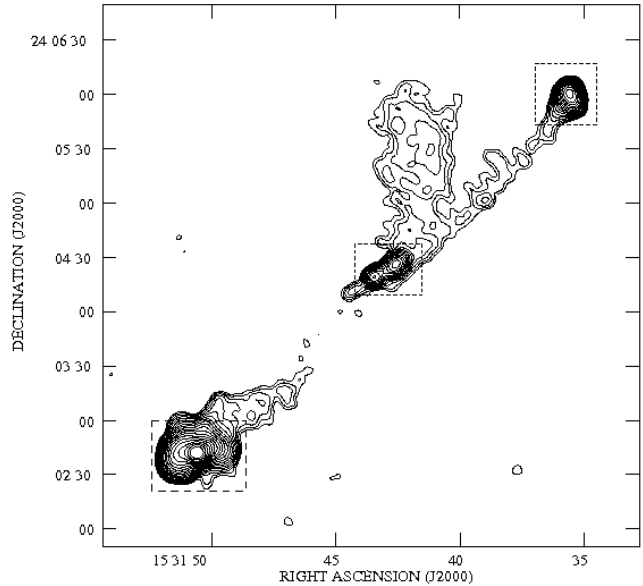}
\caption{1.5-GHz VLA A+B+C array image at $6''$ resolution of the
  entire extent of 3C\,321. Prominent radio emission is detected from
  the core, inner jet, and hotspots, together with a transverse extension to the radio lobe that lies to
  the north. The dashed boxes mark the approximate regions we used to
  image in detail the inner jet and hotspots in subsequent figures.
  Contours at $0.8 \times (1,2,4\dots)$ mJy beam$^{-1}$}
\label{vla_whole}
\end{center}
\end{figure}

\begin{figure}
\begin{center}
\includegraphics[angle=0,width=17cm]{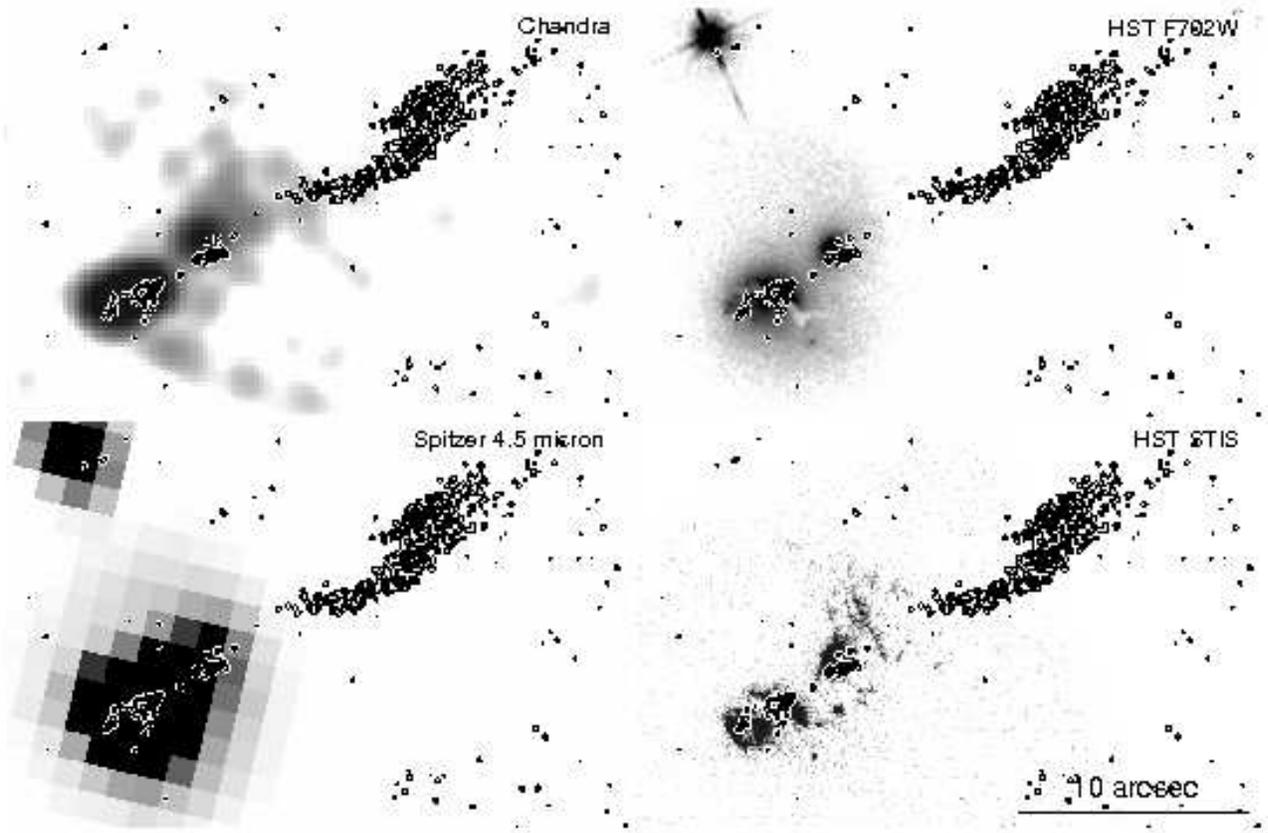}
\caption{Multiwavelength images of the nucleus, companion, and inner jet. Clockwise from top left: 0.5--2 keV {\it Chandra}, {\it HST} F702W, {\it HST} STIS NUV, and {\it Spitzer} IRAC 3.6 $\mu$m.  Overlaid on all images are contours from the combined MERLIN+VLA 1.5 GHz radio data. The \Ch image has been rebinned to 0.246$''$~pixel$^{-1}$ and subsequently smoothed with a Gaussian of $\sigma$=2 pixels.}
\label{fourpanel_center}
\end{center}
\end{figure}

\begin{figure}
\begin{center}
\includegraphics[angle=0,width=15cm]{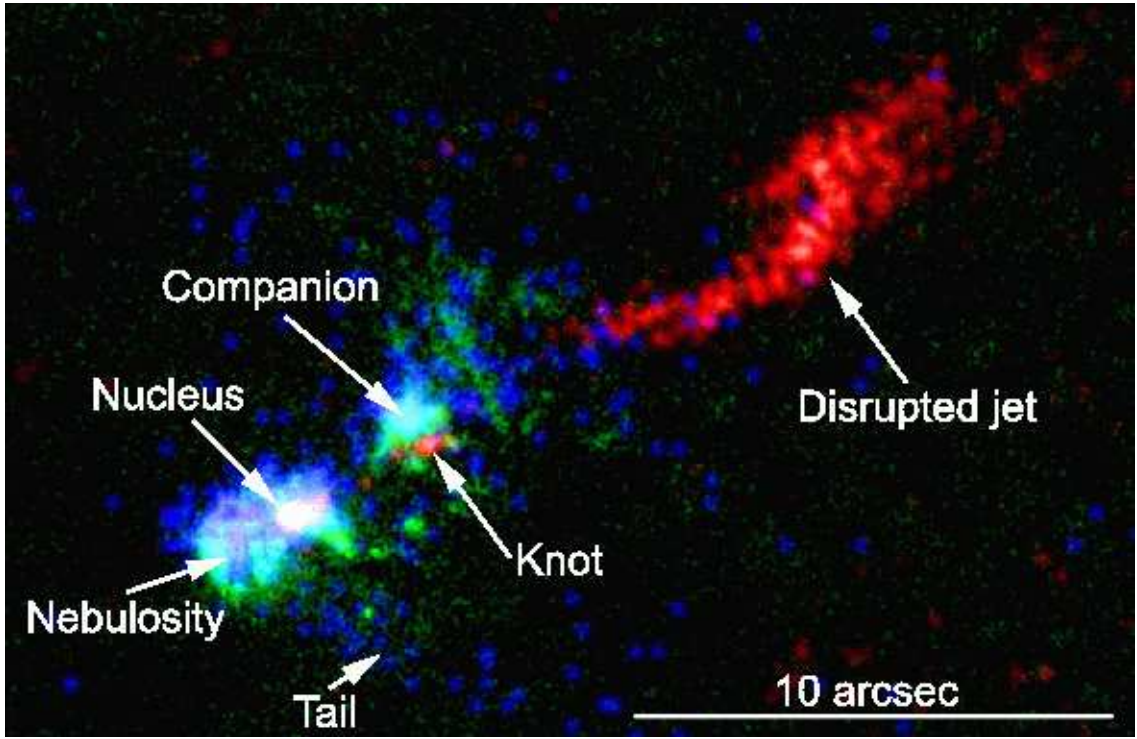}
\caption{Multiwavelength overlay of the nucleus, companion, and inner jet, using data from MERLIN+VLA 1.5 GHz radio ({\it red}), {\it HST} STIS NUV ({\it green}), and {\it Chandra} 0.5--2 keV ({\it blue}) observations.}
\label{multiwavelengthoverlay}
\end{center}
\end{figure}

Figure~\ref{vla_whole} shows a 4.8-GHz VLA A+B+C array image of 3C\,321, smoothed with a Gaussian of FWHM 6$''$, which illustrates the complex overall morphology of 3C\,321. A compact radio core and two hotspots are observed, together with a one-sided `inner jet'. A transverse extension to the radio lobe is also observed to the north. Figure~\ref{fourpanel_center} shows {\it Chandra} 0.5--2 keV, {\it HST} F702W, {\it HST} STIS NUV, and {\it Spitzer} IRAC 3.6~$\mu$m images of the inner $\sim$20$''$ of the source, with contours from the combined MERLIN+VLA 1.5 GHz radio observations overlaid. Figure~\ref{multiwavelengthoverlay} shows an overlay of the MERLIN+VLA, {\it HST} STIS, and {\it Chandra} data. 

Both Figures~\ref{fourpanel_center} and~\ref{multiwavelengthoverlay} show that an arcsecond-scale jet emerges from the compact radio core and points towards the northwest. A knot of radio emission is seen at a distance of 3.5$''$ (6.2 kpc) northwest of the core, and lies immediately to the south of the companion galaxy, but is not spatially coincident with it. The knot is resolved along the jet axis, and marginally resolved (FWHM=0.18$''$ from fitting to the MERLIN data) in the transverse direction. At a distance of $\sim$13~kpc from the core, there is a noticeable brightening of the radio jet, after which it appears to dramatically broaden and bend. The width of the diffuse jet at its widest point is 2.3$''$ (4.1~kpc). This broadened radio jet extends about 30~kpc from the jet knot. No X-ray emission associated with the broadened jet is detected. The 1.4- to 4.8-GHz spectral index of both the knot and diffuse jet are $\alpha$$\sim$0.55$\pm$0.05, as measured from maps made from data with matching shortest baselines. The uncertainties are a combination of random errors on the flux measurement estimated from the off-source noise and systematic errors in the absolute flux density calibration.

\begin{figure}
\begin{center}
\includegraphics[angle=0,width=15cm]{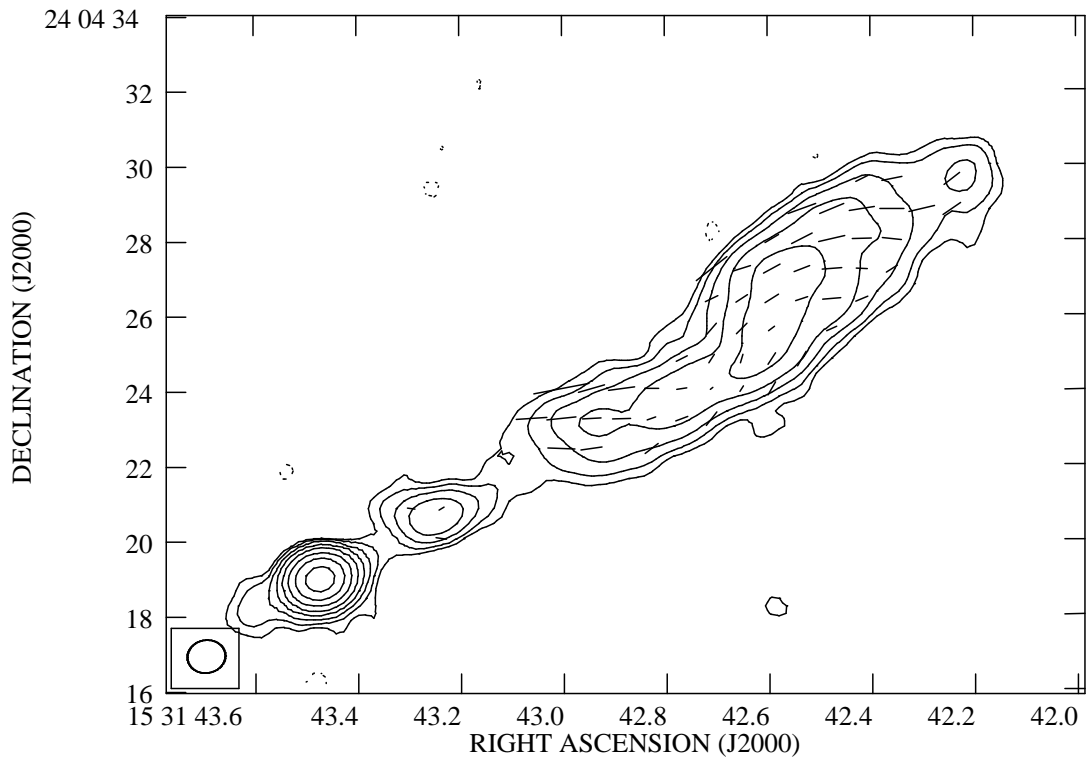}
\caption{5-GHz A+B+C VLA polarization map of the inner $\sim$20$''$ of 3C\,321. The data have a resolution of $1.02\times$0.88 arcsec, and 100\% polarization is indicated by a vector length of 1.6 arcsec. The position angle of the vectors is 90 degrees rotated from the {\bf E}-vector.}
\label{polarizationmap}
\end{center}
\end{figure}

In Figure~\ref{polarizationmap}, we present a polarization map of the
inner regions of the jet, based on 5-GHz A+B+C VLA data with a
resolution of 1.02$\times$0.88$''$. Here, the position angle of the
vectors is 90$^\circ$ rotated from the {\bf E}-vector, so that they
indicate the (approximate) direction of the magnetic field. The knot
just south of the companion is the first place where significant
polarization is evident, and it appears approximately parallel to the the jet direction. In the diffuse jet itself, all the magnetic field vectors tend to broadly lie in the same sense, along the jet.

\subsection{The Large-Scale Jet}
\label{jet-xray}

\begin{figure}
\begin{center}
\includegraphics[angle=360,width=16cm]{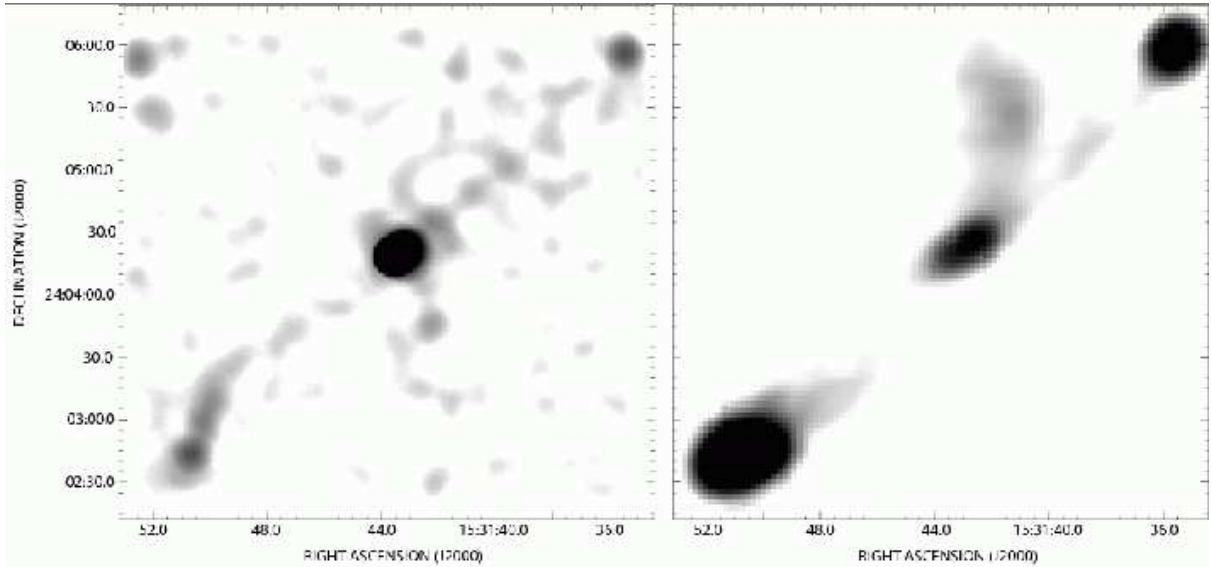}
\caption{{\it (left):} 0.5--2 keV \Ch of 3C\,321, convolved with a Gaussian of $\sigma$=10$''$, with unrelated point sources in the vicinity of the nucleus and hotspots removed. {\it (right):} 1.5-GHz VLA A+B+C array grayscale image of 3C\,321. The images show that the southeast X-ray jet bends an apparent angle $\sim$40$^{\circ}$ before it enters the southern hotspot. X-ray emission from the northern hotspot is also observed.}
\label{chandra_counterjet_smooth}
\end{center}
\end{figure}

Figure~\ref{chandra_counterjet_smooth} shows both a \Ch 0.5--2 keV image and  4.8-GHz VLA A+B+C array image of the entire extent of 3C\,321. The southeast jet is observed to bend an angle $\sim40^{\circ}$ at a distance 1.63$'$ (174 kpc) from the compact core before it terminates at the southern hotspot. On the west side of the nuclear region, there is an enhancement of X-ray emission between the nucleus and hotspot, which is significant at $\sim3\sigma$, while on the East side of the nucleus there is a more significant ($5\sigma$) detection of X-ray emission.

We extracted the X-ray spectrum of the southeast jet as it enters the hotspot using the {\sc CIAO} routine {\sc specextract}. The spectrum was extracted from a $14''\times30''$ rectangle, with background sampled from an adjacent rectangle free from point sources. There are $\sim30$ background-subtracted counts in the extraction region. The spectra were grouped to 15 counts per bin. We modeled the 0.5--7~keV spectrum with a single power law of photon index frozen at 2, modified by Galactic absorption, and found the unabsorbed 1-keV flux density of the southeast X-ray jet to be $0.69\pm0.28$ nJy. In addition, we used 1.5- and 4.8-GHz VLA data to measure the radio flux densities in an identical region, and found them to be 17.1~mJy and 7.1~mJy, respectively. This gives a radio-to-X-ray spectral index, $\alpha_{\rm RX}\sim0.9$.

\subsection{Hotspots}
\label{hotspot-xray}

\begin{figure}
\begin{center}
\includegraphics[angle=360,width=8cm]{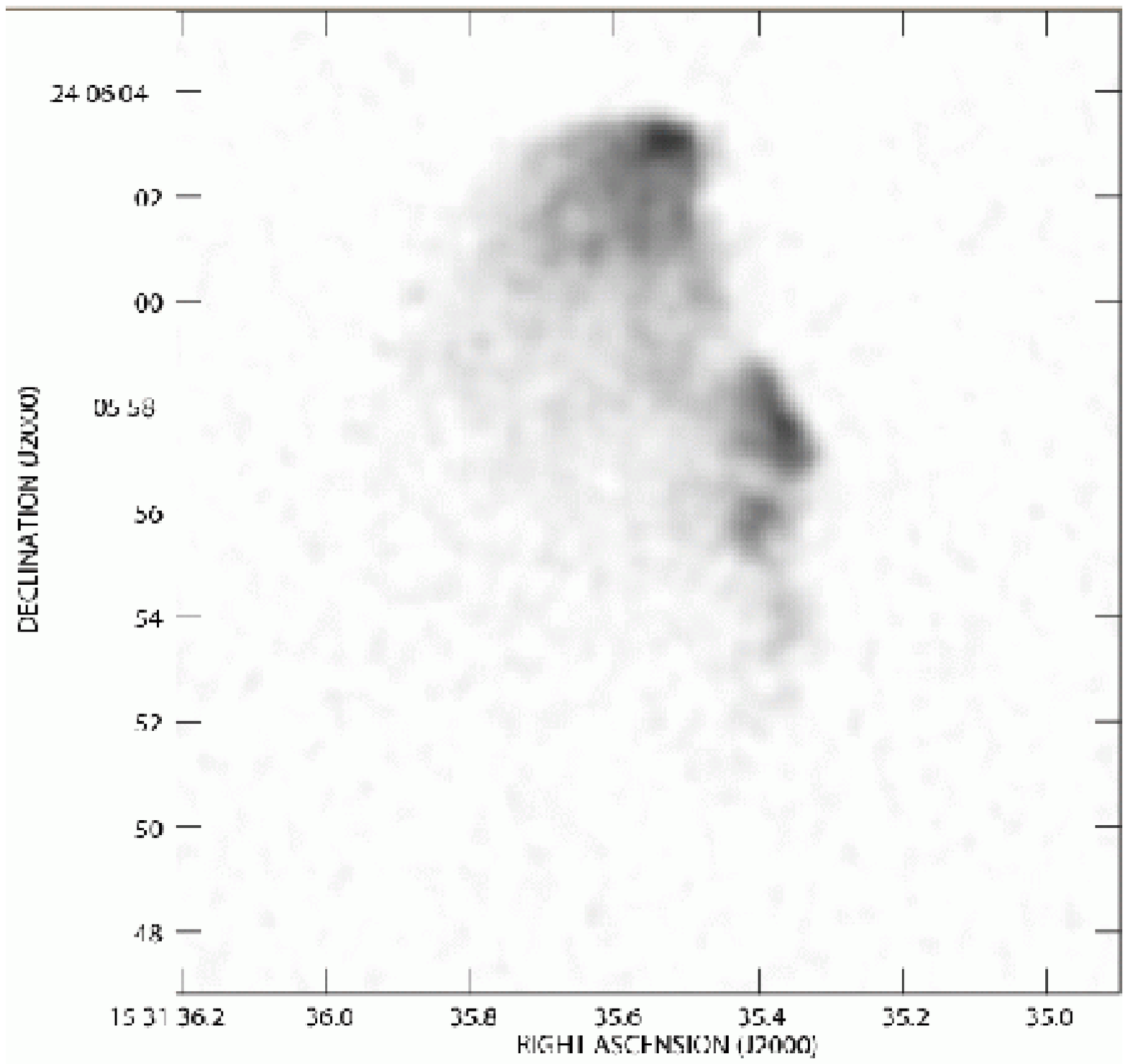}
\includegraphics[angle=360,width=8cm]{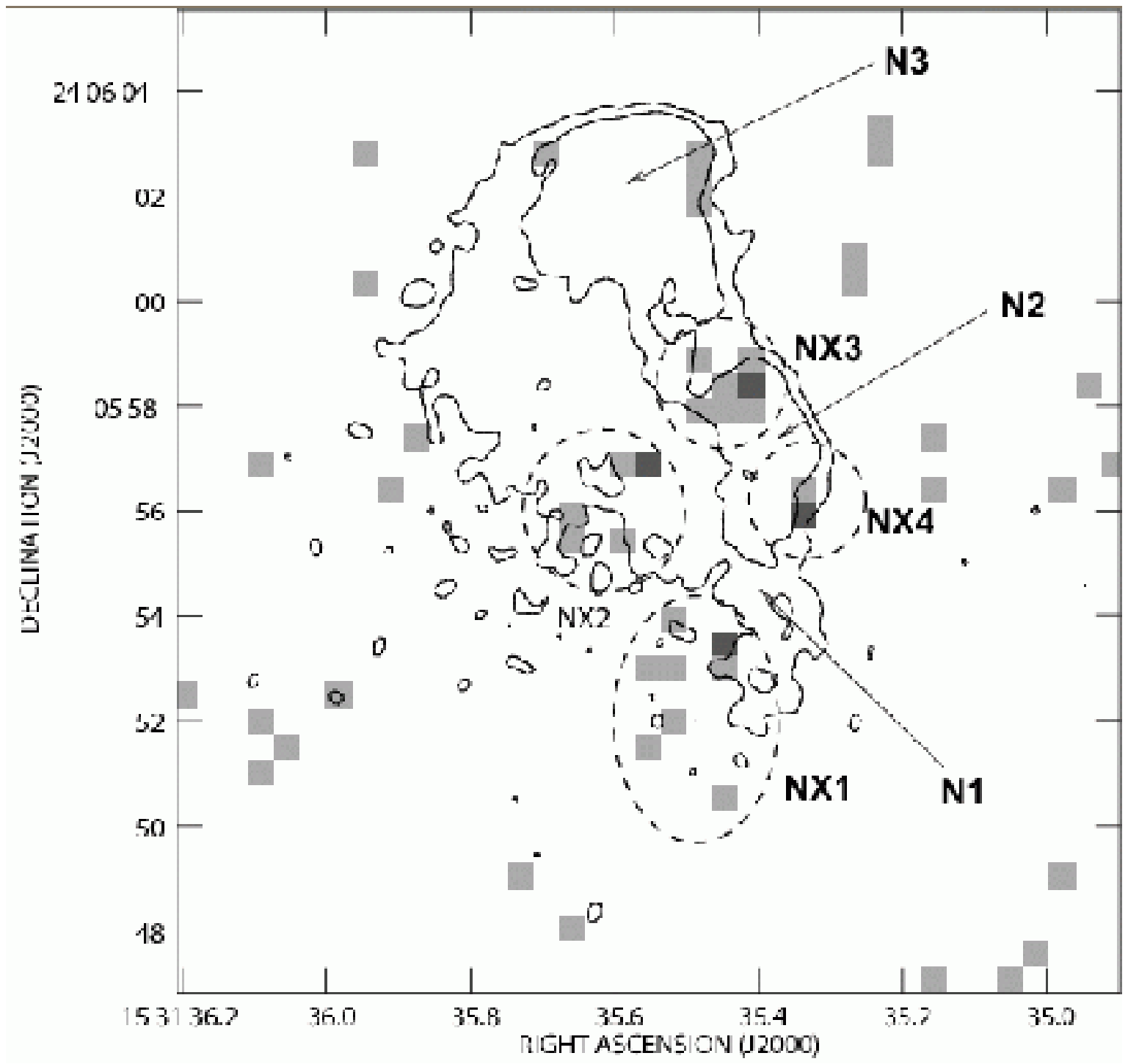}
\caption{The northern hotspot of 3C\,321. {\it (left):} 4.8-GHz VLA grayscale image with $0\farcs46$ resolution: black is 3 mJy beam$^{-1}$. {\it (right):} 0.5--5 keV \Ch X-ray image (binned in standard $0\farcs492$ pixels), with contours from the 4.8-GHz VLA data at $0.2 \times (1,4,16\dots)$ mJy beam$^{-1}$ overlaid. Regions marking the prominent radio features (N1, N2, N3) and X-ray features (NX1, NX2, NX3, NX4) are also shown.}
\label{chandra_nhotspot_radiocontours}
\end{center}
\end{figure}

\begin{figure}
\begin{center}
\includegraphics[angle=360,width=8cm]{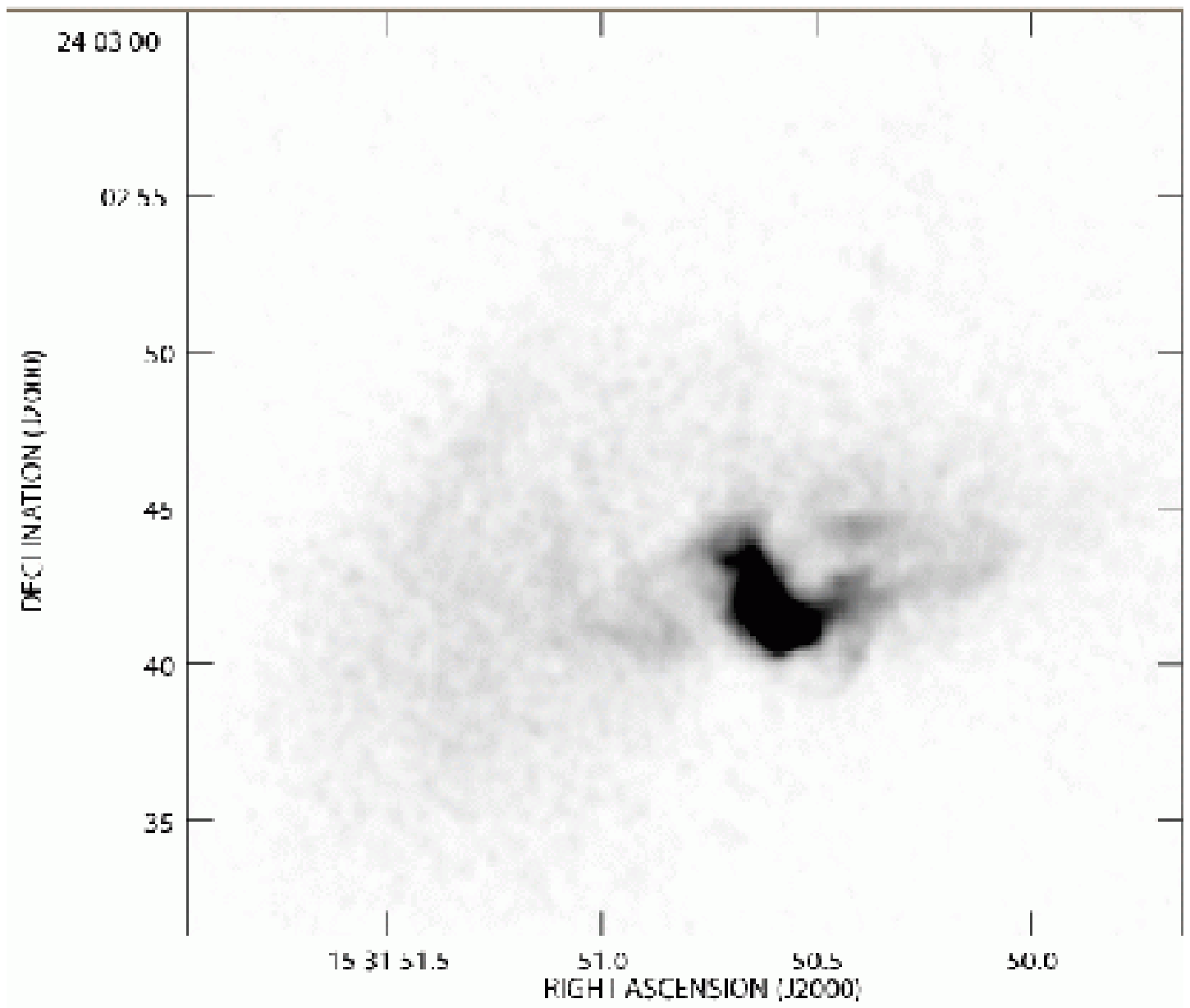}
\includegraphics[angle=360,width=8cm]{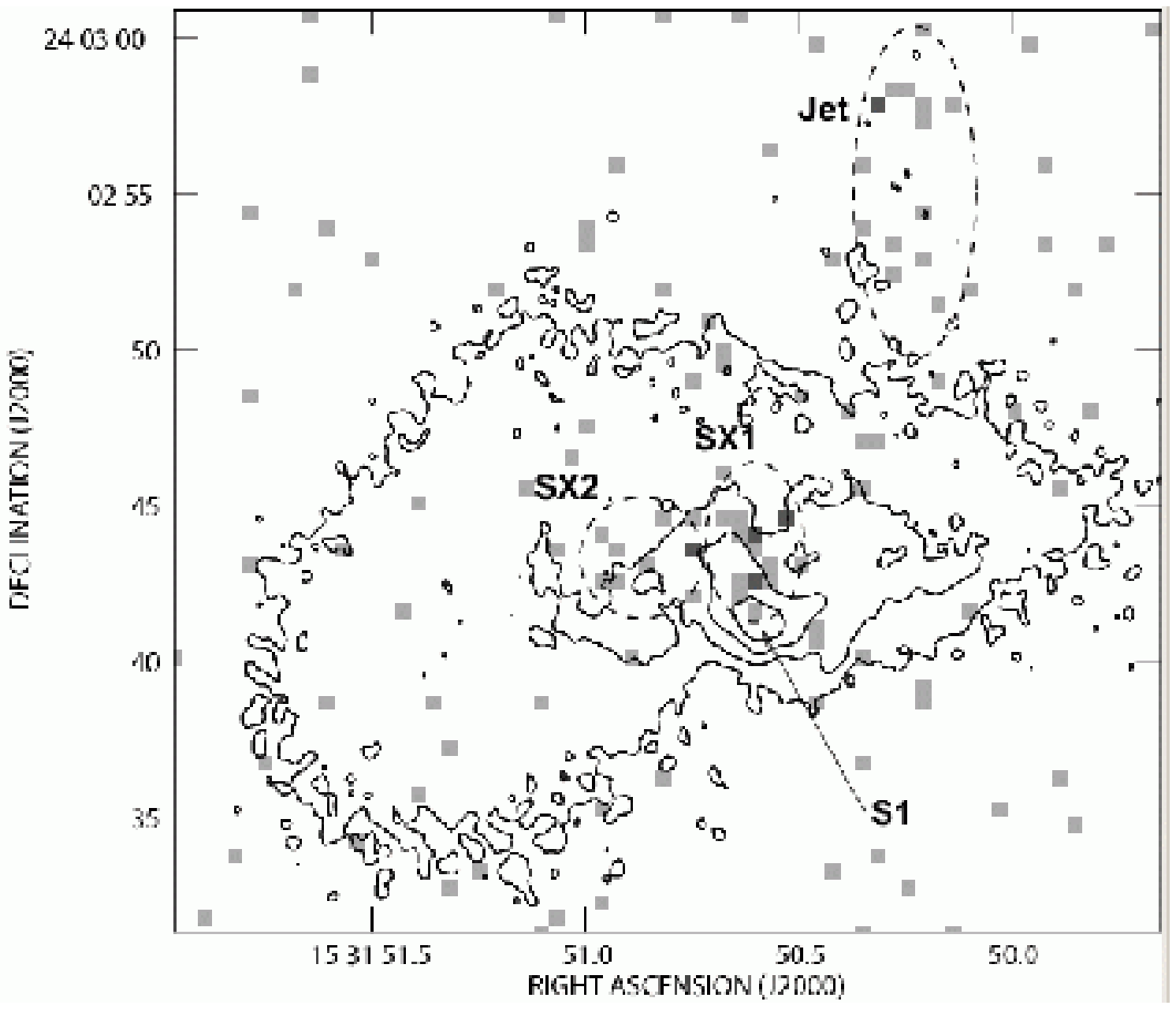}
\caption{The southern hotspot of 3C\,321. {\it (left):} 4.8-GHz VLA grayscale image  with $0\farcs46$ resolution: black is 5 mJy beam$^{-1}$. {\it (right):} 0.5--5 keV \Ch X-ray image (binned in standard $0\farcs492$ pixels), with contours from the 4.8-GHz VLA data at $0.2 \times (1,4,16\dots)$ mJy beam$^{-1}$ overlaid. Regions marking the prominent radio feature (S1) and X-ray features (SX1, SX2) are also shown.}
\label{chandra_shotspot_radiocontours}
\end{center}
\end{figure}

Both hotspots of 3C\,321 are detected in the X-ray.
Figures~\ref{chandra_nhotspot_radiocontours}
and~\ref{chandra_shotspot_radiocontours} show images of the radio and
X-ray emission in the northern and southern hotspots of 3C\,321,
respectively. In the northern hotspot, radio emission is detected in
three prominent regions, which we denote N1, N2, and N3 in
Figure~\ref{chandra_nhotspot_radiocontours}. These features are
accompanied by more diffuse radio emission that extends to larger
scales, a tail of radio emission to the feature N1, and possibly a
filament of emission that links N2 and N3. In the X-ray, we identify
four distinct regions of emission, which we denote NX1, NX2, NX3, and
NX4. None of these features is spatially coincident with the brightest
parts of the radio features N1, N2, or N3: both NX1 and NX2 lie
$\sim$2.5--3$''$ ($\sim$4.5--5.5~kpc) away from N1 in the direction of
the nucleus, while NX3 and NX4 lie towards the edges of N2. The 0.5--2
keV X-ray luminosity of the northern hotspot is
$\sim4\times10^{40}$~ergs~s$^{-1}$. No infrared emission associated
with the northern hotspot is detected with {\it Spitzer}.

Both radio and X-ray emission are detected from the southern jet as it enters the hotspot, as indicated in Figure~\ref{chandra_shotspot_radiocontours}. There is only one compact region of radio emission (marked S1 in Figure~\ref{chandra_shotspot_radiocontours}) in the southern hotspot, which is accompanied by a small tail just to its north, together with a complex distribution of diffuse and filamentary emission to the east and west. There is no X-ray detection of a compact component corresponding to S1; instead, we observe an elongated region of X-ray emission (labeled SX1) that lies $\sim$2.8$''$ ($\sim$5~kpc) north of S1 and extends along a similar position angle to it. In addition, there is evidence for X-ray emission (labeled SX2) that lies $\sim$3.8$''$ ($\sim$6.5~kpc) NE of S1. The 0.5--2 keV X-ray luminosity of the southern hotspot is $\sim4\times10^{40}$~ergs~s$^{-1}$. No infrared emission is detected with {\it Spitzer} in the southern hotspot.

\subsection{The X-ray spectra of the Host and Companion Galaxies}
\label{spectra-xray}

\begin{figure}
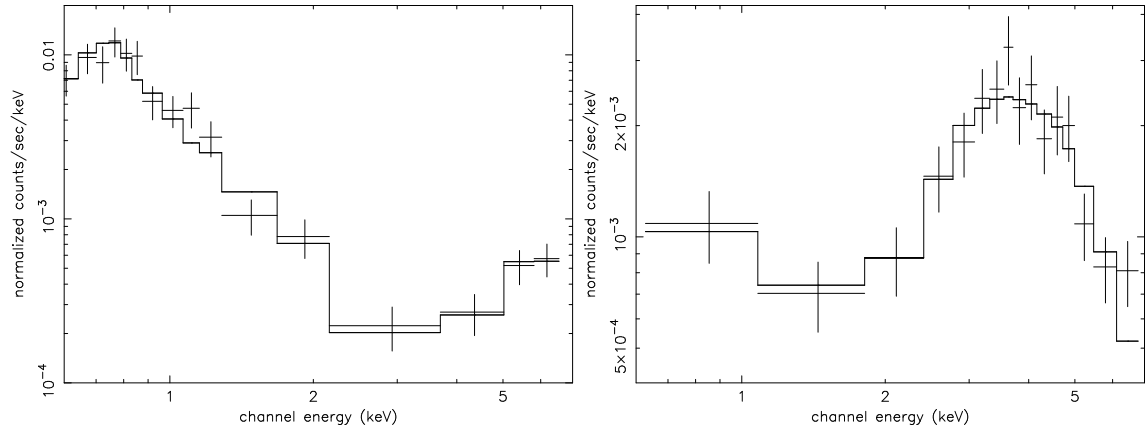

\begin{center}
\includegraphics[angle=270,width=7.5cm]{f8a.eps}
\includegraphics[angle=270,width=7.5cm]{f8b.eps}
\caption{(a) 0.5--7 keV Chandra X-ray spectrum of the core. The model fit shown is the sum of an absorbed power law, a second unabsorbed power law, and a collisionally ionized plasma. (b) 0.5--7 keV Chandra X-ray spectrum of the companion. The model fit shown is the sum of an absorbed power law, and a second unabsorbed power law.}
\label{core+companionspectrum}
\end{center}
\end{figure}

We extracted spectra from the nucleus of 3C\,321 using the {\sc CIAO} routine {\sc psextract}.  The spectra were extracted from a source-centered region of radius $1.23''$, with background sampled from an annulus surrounding the northwestern portion of the source region, of inner radius $1.23''$ and outer radius $2.46''$. The spectra were grouped to 20 counts per bin. The spectrum of the nucleus is shown in Figure~\ref{core+companionspectrum}a. Our best-fitting spectral model is the sum of a heavily absorbed [$N_{\rm H}= (1.04^{+0.55}_{-0.21})\times10^{24}$ cm$^{-2}$] power law of photon index frozen at 1.7, a second unabsorbed power law of photon index frozen at 2, and a collisionally ionized ({\sc apec}) plasma of temperature $kT=0.49^{+0.15}_{-0.18}$ keV and abundance fixed at solar. We decided to freeze the photon indices of the power laws at their canonical values (\citealt{evans06}), although the fit is relatively insensitive to this choice owing to the limited number of counts. The unabsorbed 0.5--10 keV luminosity of the heavily absorbed power law is $(4.3^{+8.8}_{-2.6})\times10^{43}$ ergs s$^{-1}$. The parameters of the best fitting model are given in Table~\ref{extensionspectrum_s_table}.

In addition, we extracted the spectrum of the companion galaxy, using a circular source region of radius 1.23$''$. Background was sampled from two surrounding pie slices of inner radius 1.23$''$ and outer radius 2.66$''$, between the position angles 76$^{\circ}$ to 187$^{\circ}$, and 260$^{\circ}$ to 358$^{\circ}$, respectively, in order to mask out any contributions from the known small-scale extended X-ray emission. The spectrum is shown in Figure~\ref{core+companionspectrum}b, and clearly indicates the presence of an absorbed component. We modeled the spectrum with the combination of an absorbed [$N_{\rm H}=(1.01^{+0.46}_{-0.35})\times10^{23}$~cm$^{-2}$] power law of photon index $\Gamma=1.83^{+0.25}_{-0.31}$, and a second component (either an unabsorbed power law, or a collisionally ionized plasma, though there are insufficient counts at low energies to distinguish between the two). The unabsorbed 0.5--10 keV luminosity of the heavily obscured power law is $(1.4^{+1.1}_{-0.7})\times10^{43}$ ergs s$^{-1}$.

\subsection{Diffuse Gas in the Vicinity of the Nuclear System}
\label{diffuse_and_tails}

Figures~\ref{fourpanel_center} and~\ref{multiwavelengthoverlay} illustrate the complex morphology of circum-source gas associated with the host and companion galaxies. The {\it HST} STIS NUV and [O~{\sc iii}] images show a series of knots in a ring-like nebulosity that extend $\sim$2.2$''$ ($\sim$4~kpc) southeast of the host-galaxy nucleus, together with arcs and filaments that extend in a fan-like shape $\sim$4.4$''$ ($\sim$8~kpc) northwest of the companion galaxy, as discussed by \cite{hur99} and \cite{allen02}. There is extended X-ray emission southeast of the nucleus, which is spatially coincident with the NUV and [O~{\sc iii}] nebulosity, although the limited angular resolution of \Ch prevents us from examining the X-ray morphology of these features in detail. There is an excess of X-ray emission northwest of the companion galaxy, which has a similar spatial distribution to the NUV and [O~{\sc iii}] emission.

\begin{figure}
\begin{center}
\includegraphics[angle=0,width=16cm]{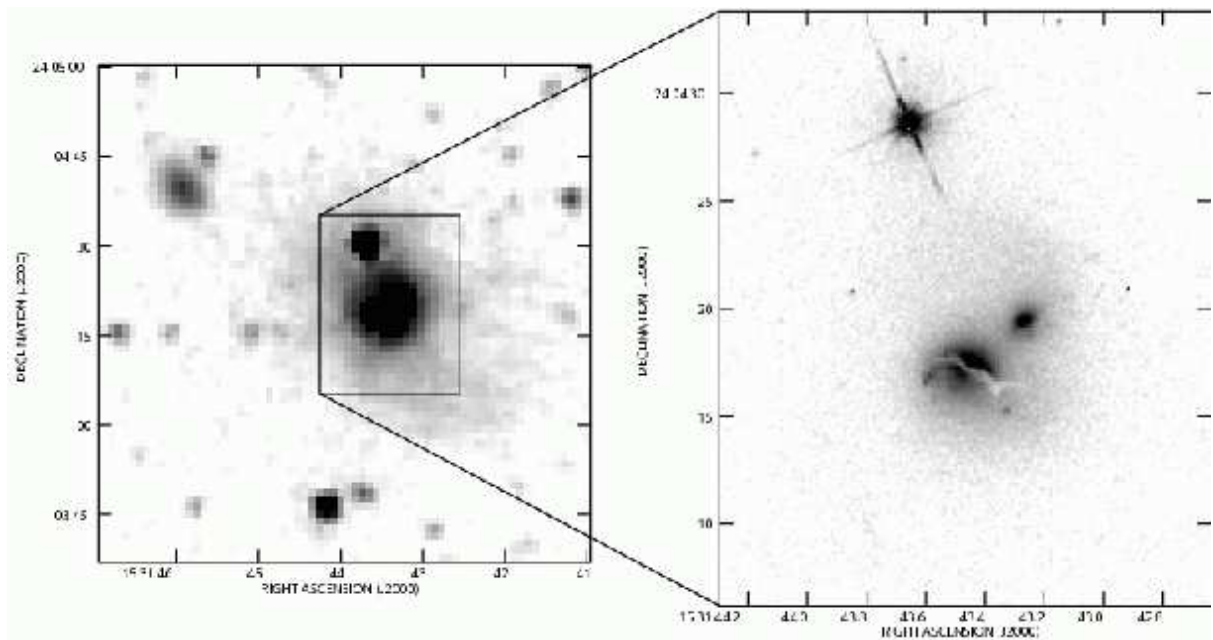}
\caption{{\it Spitzer} IRAC 4.5~$\mu$m {\it (left)} and {\it HST} WFPC2 F702W {\it (right)} images of the common gas envelope of the host galaxy of 3C\,321 and the companion galaxy.}
\label{spitzer_hst_tidal_tail}
\end{center}
\end{figure}

On larger scales, {\it Spitzer} IRAC 4.5~$\mu$m and {\it HST} F702W images show a tail that extends south of the host galaxy of 3C\,321 before curving to the SW, with a total extent of $\sim$30$''$ ($\sim$55~kpc) (see Fig.~\ref{spitzer_hst_tidal_tail}). X-ray emission is seen along its northern edge (Figs.~\ref{fourpanel_center} and~\ref{multiwavelengthoverlay}) and extends $\sim$10$''$ ($\sim$18~kpc). The X-ray spectrum of this feature is well-described by a {\sc apec} model of temperature $kT=0.31^{+0.11}_{-0.06}$~keV and solar abundance, and has an unabsorbed  0.5--2 keV luminosity of $(5.2\pm1.9)\times10^{40}$~ergs~s$^{-1}$. A fainter tail pointing to the north and curving to the east is also observed in Figure~\ref{spitzer_hst_tidal_tail}.

\section{Interpretation}
\label{interp}

The results presented in this paper demonstrate that the morphology of 3C\,321 is highly atypical for a standard FRII radio galaxy. Its one-sided, diffuse $\sim$30~kpc jet is similar to an FRI-type jet; yet the hotspots observed on much larger ($\sim$250~kpc) scales are compact and exhibit on-going particle acceleration to X-ray energies. In this section, we explore the physical origin of these features, and argue that an interaction between the jet and the companion galaxy is the most likely explanation for its flared nature (\S\ref{interp-evidence}). We suggest that this interaction is a relatively recent one (\S\ref{interp-timescales}), given that {\it in situ} particle acceleration is still taking place in the hotspots. We provide some simple example models in \S\ref{interp-models} that demonstrate that a jet--companion interaction is physically feasible, and show that the situation of two merging galaxies, each of which hosts AGN, is a rare one (\S\ref{interp-pair}). We also show that the observed emission from the hotspots and large-scale X-ray jet is consistent with that seen in other FRII galaxies (\S\ref{interp-hotspots}).

\subsection{Evidence for a jet--companion interaction?}
\label{interp-evidence}

We begin by considering the physical origin(s) of the jet knot at 6.2 kpc from the core and the subsequent flaring of the jet out to $\sim$30 kpc. The diffuse jet morphologically at least resembles an FRI-like structure. However, the magnetic field direction in 3C\,321 very broadly lies in the jet direction and so appears {\it not} to behave like FRI jets. In standard, twin-jetted FRI sources [``weak-flavor jets'' in the nomenclature of \cite{lai93} and references therein] the $\sim$kpc-scale magnetic field direction is usually transverse to the jet direction in the center, though sometimes a longitudinal field is observed towards the edge (e.g., \citealt{har96,lai06,dul07}). FRII jets, WAT jets (e.g.,~\citealt{eil84}), or limb-brightened FRIs [all ``strong-flavor jets'' as described by Laing et al.] are those with a predominantly parallel magnetic field.

Given the close spatial proximity of the jet knot to the
companion galaxy, at least in projection, it is immediately tempting
to suggest that the jet is being disrupted by an interaction between
it and the companion. There is little velocity difference between the
host galaxy of 3C\,321 and the companion, while the two-sided
large-scale radio morphology and constraints on the beaming angle of
the X-ray jet (\ref{interp-hotspots}) are consistent with the idea
that the jet is close to the plane of the sky. Therefore, the {\it a priori} probability
that the jet will see some part of the companion galaxy mass/gas
distribution is quite high. We emphasize that any interaction between the jet and the companion
galaxy cannot be entirely catastrophic: the jet continues to propagate after the interaction,
albeit in a somewhat disrupted manner. If the jet termination were
complete, the radio `knot' at the point of the interaction would
essentially be a radio hotspot, in which all the bulk kinetic power of
the jet would be dissipated, and we might expect a much higher radio
luminosity (the northwest hotspot is two orders of magnitude brighter)
together with associated (radio lobe-like) backflow.

We caution that we cannot rule out the fact that this apparent interaction is merely a projection effect, and that the observed flared morphology of the jet is instead an intrinsic property of the outflow. In the absence of detailed imaging of the radio jet to either side of the knot, it is not possible completely to disprove that the change of jet direction is caused, for example, by a Kelvin-Helmholtz type instability (e.g., \citealt{har79}). A single jet bend is not possible, since the appearance of the jet should be that of a projected helix in the Kelvin-Helmholtz framework: in other words, some distance back from the bend there would have to be a bend in the opposite sense. This problem could be overcome if the growth length of the instability is short, so that the previous bend could be quite small, but then suppressing the other modes of K-H instability would be difficult. We also cannot rule out the possibility that the jet is intrinsically broad to begin with; i.e., that the `knot' is in fact an edge-brightened feature of a lower surface brightness structure that extends further north. However, given that the compact components seen in the northwest hotspot are substantially smaller than the diffuse jet at its widest point (2.3$''$=4.1~kpc), it is difficult to see how the jet could terminate in such compact features if the hotspots are the beamheads of active jets. Although these problems are not insurmountable, we consider the {\it simplest} interpretation to be that the jet interacts with the companion galaxy, and we explore the consequences of this explanation in the following subsections.

\subsection{Timescales of the interaction}
\label{interp-timescales}

The northwest hotspot has compact components in the radio and emits X-ray synchrotron radiation (\citealt{har04}) that is indicative of on-going particle acceleration. Energy is therefore required to sustain the hotspots in X-rays, and interpreting the hotspots as the beamheads of active jets is a logical way to maintain the energy. This suggests that the jet--companion interaction is {\it not} a steady-state situation. We argue that (1) the jet has been disrupted on a timescale
less than the jet travel time to the end of the lobe --- probably
comparable to the light travel time of $\sim 10^6$ years, assuming
that the source is close to the plane of the sky, and that (2) the jet
flow to the northwest hotspot will only be disrupted for as long as the
jet--companion interaction takes place. From fitting to the MERLIN
data, we estimate that the width of the jet close to the bright knot
is $\sim$0.4 kpc, and we assume that this represents the width of the
energy-carrying channel. If the companion galaxy is moving at a speed
of $\sim 200$ km~s$^{-1}$, which is plausible given the range of
radial velocities observed in the optical line emission
(\citealt{rob00}), then the time taken for gas associated with the companion to intercept the entire
width of the jet is $2 \times 10^6$ years. If the interaction has
lasted for less than $10^6$ years, then the projected length of the
disrupted jet ($\sim$30~kpc) implies that the speed of the jet
downstream of the interaction is $\ga 0.1c$. This is not implausible
given that we expect the pre-shock bulk speed of the jet to have been
relativistic.

\subsection{Possible disruption models}
\label{interp-models}

We presented evidence that an interaction between the jet and the companion galaxy is qualitatively the most simple explanation for the observed broadening in the jet, and that the interaction is not a catastrophic one for the continued propagation of the jet. Here we discuss two simple models that demonstrate that the model of a jet--environment interaction is at least physically plausible, but we do not aim to distinguish between them.

One possible model is that the jet is gently disrupted by an interaction with the gaseous ISM of the companion galaxy. For jets with similar powers to 3C\,321, interactions with a diffuse medium (e.g., \citealt{sok88,wang00}), plausibly of lower density than gas associated with the companion galaxy in 3C\,321, are seen to be capable of producing sharp bends (e.g., \citealt{eil84,lok95,har05}). In the case of 3C\,321, the jet is observed to bend at most several tens of degrees, and so this type of interaction seems plausible.

Another model is that the jet has encountered a clump of dense cold material which causes an internal shock in the jet and creates a deflection in the outflow. The flat spectral index ($\alpha$$\sim$0.55) of the knot between 1.5 and 4.8 GHz is not inconsistent with some particle acceleration having taken place. For example, $\alpha$=1 would be difficult to reconcile with a model in which the knot is a physical manifestation of a strong interaction of the jet with its environment. \cite{choi07} performed relativistic hydrodynamical simulations of such an interaction, and found that a knot of synchrotron emission is observed at the interaction point, similar to the situation in 3C\,321. They found that the jet can deflect through a significant angle in direct response to its interaction with the cloud but remain stable afterward, depending on the density contrast of the cloud to the ambient medium. In 3C\,321, the observations can be explained if the shock in the jet also translates bulk kinetic energy into internal energy of the jet, causing the jet to expand downstream of the shock, and giving rise to the region immediately downstream of the knot with little or no detected radio emission. The jet would then recollimate and brighten, perhaps at a reconfinement shock, creating the bright diffuse radio emission we see at 13 kpc from the nucleus. For consistency with this description, we require that the external obstacle should not have been accelerated to relativistic speeds by the thrust of the jet over the timescale of the interaction: that is, $p_{\rm jet} A_{obs} t/m_{obs} \la 0.1c$, where $p_{\rm jet}(=P_{\rm jet}/A_{\rm obs}c)$, $A_{\rm obs}$ and $m_{\rm obs}$ are the pressure of the jet, and cross-sectional area and mass of the obstacle, respectively. Even in the limiting case that all the thrust of the jet acts to accelerate the obstacle over $t = 10^6$ years, though, this only requires $m_{obs} \ga 10^5 M_\odot$, in the mass range for giant molecular clouds in our own Galaxy. 

There is at present no observational way of
constraining these models. The existing {\it Chandra} observations can
only provide very approximate constraints on the density of hot gas in the
region of the companion, while dense clumps of cold gas are not
detectable by any available observation. In addition, the kinetic
power of the jet is extremely uncertain. Our object in presenting
these models is simply to show that such jet-environment interactions
are physically possible. Detailed hydrodynamical simulations with the
aim of reproducing the observed radio structure of the jet might
enable us to make some progress in this area.

\subsection{The nature of the galaxy pair}
\label{interp-pair}

The extended gas envelope seen in the {\it HST}, {\it Spitzer} and {\it Chandra} observations (Section~\ref{diffuse_and_tails}) suggests that we are seeing a merger between the host galaxy of 3C\,321 and the smaller (companion) galaxy, and that the companion galaxy is orbiting the center of mass in the clockwise direction. The temperature of the south X-ray gas `tail' is $0.31^{+0.11}_{-0.06}$~keV, which suggests that it was previously associated with the galaxy-scale atmosphere of either the host or companion galaxy. 

Our detection of heavily absorbed nuclear X-ray emission in the companion galaxy of 3C\,321 (\S\ref{spectra-xray}), together with its optical emission-line spectrum (\citealt{fil87}), circumnuclear optical polarization (\citealt{dra93}), and broad H$\alpha$ wings (\citealt{rob00}) confirm that the companion galaxy is an AGN in its own right. No radio emission is associated with the nucleus of the companion galaxy, leading to its classification as a Seyfert 2 galaxy. The situation of two merging galaxies, each of which host AGN, is a rare one, at least for low redshift systems for which good-quality X-ray data exist. The best-studied example of a binary AGN is in the ultraluminous infrared galaxy NGC\,6240 (\citealt{kom03}), which shows neutral Fe~K$\alpha$ line emission in both nuclei. There is a handful of other examples of possible binary AGN, but in these cases the second galaxy tends to be significantly less luminous than its partner so that AGN emission cannot be unambiguously distinguished from starburst activity (e.g., \citealt{bal04,jim07}).

In 3C\,321 we see evidence for two heavily obscured AGN, each with luminosities in excess of $10^{43}$ ergs~$^{-1}$, {\it and} an FRII jet-gas interaction. As these are unusual, it is worth considering whether there can be any {\it causal} link between the two. One highly speculative possibility is that an interaction with the radio galaxy jet helped to drive material onto the nucleus of the companion, triggering its currently observed AGN activity. However, as the nucleus is currently 0.85 kpc to the north of the peak of the jet, any such interaction must have taken place $\sim 4$ Myr ago for a nominal galaxy speed of 200 km~s$^{-1}$: thus, by the arguments used above, it cannot have been the same as the interaction that is currently taking place, since otherwise no compact hotspots would be seen. In this picture we would require two interactions with a re-establishment of the well-collimated jet flow in between, so even here we have to be `lucky' to see both an AGN in the companion and an interaction between the jet and companion-associated gas. An alternative interpretation is that the merger triggered both AGN {\it independently}, in which case several of these problems would be removed.

\subsection{Hotspots, jets and emission mechanisms}
\label{interp-hotspots}

3C\,321, like many other FRII radio sources (e.g., \citealt{lea97}) has morphologically different northern and southern hotspots, each of which consist of multiple compact radio components, together with spatial offsets between the radio and X-ray emission (e.g.~\citealt{har07a}). It is conventional to describe hotspot features in terms of a compact `primary' hotspot, which marks the point of termination of the jet, together with one or more diffuse `secondary' features, which may or may not be connected to the primary by some sort of collimated outflow. We follow this convention below. 

In the northern hotspot of 3C\,321, a relatively compact `primary' (N2) is observed in the radio, and is accompanied by X-ray emission (NX3 and NX4) either side of it. A possible explanation for the X-ray and radio morphology of the northern hotspot is that the jet has entered it through the `tail-like' radio structure just south of N1 (in which case the feature NX1 would represent X-ray emission from the jet) before terminating at N2. The physical reason why X-ray-emitting features could lie either side of a primary radio hotspot remains unclear, although this morphology is similar to that observed in the northern hotspot of the nearby, well-studied radio galaxy 3C~390.3 (\citealt{har07a}).

The southern hotspot is morphologically rather different from the northern hotspot. The spatial offset of $\sim1.5$$''$ (2.7~kpc) between S1 and the main component of X-ray emission (SX1) can be accounted for in several ways. For example, SX1 may {\it not} be directly related to the radio hotspot, and could for instance be synchrotron emission from i) a (deviated) jet as it enters the southern hotspot, or ii) the radio tail observed just north of S1, which could be some sort of outflow associated with S1. Alternatively, the X-ray emission may be directly related to the compact radio hotspot, but offset from it, and may either trace the location of the shock {\it now}, or be produced by inverse-Compton scattering of radio synchrotron-emitting photons in the slower (downstream) component of the flow (\citealt{gk03}). Our observations of 3C\,321 are qualitatively consistent with any of these interpretations, and so we cannot distinguish between them.

In the X-ray, the most frequently discussed emission mechanisms for
hotspots are synchrotron and synchrotron self-Compton (SSC) radiation.
\cite{har04} argued that in low-power hotspots like those of 3C\,321
the SSC process is unlikely to be important, and showed specifically
for 3C\,321 that SSC with equipartition field strengths underpredicts
the level of X-ray emission observed by several orders of magnitude.
If the hotspot X-ray emission is synchrotron, then it implies that
there is ongoing high-energy particle acceleration, and therefore
ongoing well-collimated energy input, in both hotspots. 

We see clear evidence for X-ray emission from the southern jet of
3C\,321 over $\sim$50 kpc. Jet X-ray emission has been seen in a
number of other low-power, low-redshift FRII radio galaxies
(e.g.,~\citealt{kra05,wor05}) and in those cases it seems almost
certain that the emission mechanism is synchrotron. In the case of
3C\,321 we find (using the code of \citealt{hbw98}) that the
equipartition prediction for SSC is more than 4 orders of magnitude
below the observed X-ray, while beamed inverse-Compton scattering of the CMB
requires angles to the line of sight of $\la 10^\circ$ and bulk
Lorentz factors $\ga 7$. The synchrotron process thus seems to be the
only viable mechanism to explain the detected jet X-ray emission in 3C\,321. If this is the case it requires a distributed particle acceleration process throughout the observed X-ray and radio jet. The existing optical and infrared data do not provide interesting constraints on a synchrotron model.

\section{CONCLUSIONS}

We have presented results from a radio through X-ray study of the nucleus, environment, jets, and hotspots, of the nearby FRII radio galaxy 3C\,321. Our conclusions can be summarized as follows:

\begin{enumerate}

\item The inner radio morphology of the source is atypical for
  an FRII radio source. A small-scale jet emerges from the nucleus of
  its host galaxy, produces a knot of radio emission that lies
  immediately to the south (in projection) of a smaller companion
  galaxy, and then flares and bends into a diffuse structure 35~kpc
  from the nucleus.

\item We argue that the simplest explanation for the
  morphology of the small-scale diffuse radio jet is that it is interacting with the companion
  galaxy. We have discussed two representative models, including an interaction with the ISM of the companion galaxy and a cloud of dense gas, that can account for the morphology, although we are unable to distinguish between them.

\item The northwest hotspot that lies $\gappeq$250~kpc from the core exhibits on-going {\it in situ} particle acceleration to X-ray energies, which strongly suggests that the disruption of the jet is not a steady-state situation. In other words, we argue that the jet has been disrupted on a timescale less than the light travel time to the northwest hotspot of $10^6$ years, and that the jet flow to this hotspot will only be affected for as long as the interaction takes place.

\item We cannot rule out the situation that the diffuse morphology of
the radio jet is somehow related to an intrinsic property of the jet, rather
than the extrinsic impact of the environment through which it
propagates. However, these alternative models would need additional
physics to explain, for example, why the diffuse jet appears to be
one-sided.

\item The host galaxy of 3C\,321 and the companion galaxy are in the
process of merging, and each hosts a luminous AGN. As this is an
unusual situation, we have investigated the hypotheses that the
interacting jet has driven material on to the companion galaxy,
triggering its AGN, or that the merger has triggered the AGN activity in both galaxies.

\item There are multiple sites of X-ray emission in both the northern
and southern hotspots, together with spatial offsets between the radio
and X-ray emission. We have presented several models that might
account for the radio--X-ray offsets, including a sudden bending of
the jet as it enters the hotspot, X-ray emission associated with an
outflow from the primary hotspot, and differences in the
electron-energy distributions and particle-acceleration processes in
the hotspot.

\end{enumerate}

\acknowledgements

The authors gratefully acknowledge the assistance and constructive comments of the anonymous referee, which have helped to improve this paper.
We wish to thank Paul Nulsen and Dan Harris for useful suggestions. MJH
thanks the Royal Society for a Research Fellowship. The National Radio
Astronomy Observatory is a facility of the National Science Foundation
operated under cooperative agreement by Associated Universities, Inc.
MERLIN is a National Facility operated by the University of Manchester
at Jodrell Bank Observatory on behalf of PPARC. This paper was partly
based on observations made with the NASA/ESA {\it Hubble Space
Telescope}, obtained from the data archive at the Space Telescope
Institute. STScI is operated by the Association of Universities for
Research in Astronomy, Inc. under NASA contract NAS 5-26555. This work
is based in part on observations made with the {\it Spitzer Space
Telescope}, which is operated by the Jet Propulsion Laboratory,
California Institute of Technology under a contract with NASA. This
research has made use of the NASA/IPAC Extragalactic Database (NED)
which is operated by the Jet Propulsion Laboratory, California
Institute of Technology, under contract with the National Aeronautics
and Space Administration.

\newpage
\begin{deluxetable}{lllll}
\tablecaption{Archival observations with the VLA used in this paper}
\tablehead{
Frequencies&Configuration&Date&Duration&Obs. ID \\
(GHz)      &             &    &(h)              }
\startdata
4.81,4.86&A&1986 Apr 10&3.6&AV127\\
4.81,4.86&B&1986 Aug 29&3.6&AV127\\
4.81,4.86&C&1986 Dec 02&0.3&AV127\\
1.38,1.64&A&1986 Apr 10&2.9&AV127\\
1.38,1.64&B&1986 Aug 29&2.9&AV127\\
1.38,1.64&C&1986 Dec 02&1.3&AV127
\enddata
\label{vlaobs}
\end{deluxetable}

\begin{deluxetable}{lllc}
\tablecaption{Best-fitting parameters of model fits to the X-ray spectra of the nucleus and companion galaxy}
\tablehead{
Model & Description & Parameters                & $\chi^2$/d.o.f. }
\startdata
Nucleus   & $N_{\rm H}$(PL)+PL+TH & $N_{\rm H}= (1.04^{+0.55}_{-0.21})\times10^{24}$ & 14.21/12  \\
          &                       & $\Gamma_1=1.7$ (f)&                             \\
          &                       & norm$_1$=$(3.08^{+6.21}_{-1.84})\times10^{-4}$ &      \\
          &                       & $\Gamma_2=2$ (f) &                            \\
          &                       & norm$_2$=$(3.86^{+1.08}_{-1.16})\times10^{-6}$       &                             \\
          &                       & $kT=0.49^{+0.15}_{-0.18}$ keV &       \\
          &                       & $Z=1$ (f) &                           \\
          &                       & norm$(6.67^{+3.42}_{-1.70})\times10^{-6}$  & \\
Companion & $N_{\rm H}$(PL)+PL    & $N_{\rm H}= (1.01^{+0.46}_{-0.35})\times10^{23}$ & 8.88/12                      \\
          &                       & $\Gamma_1=1.83^{+0.25}_{-0.31}$ &                            \\
          &                       & norm$_1$=$(1.14^{+0.60}_{-0.54})\times10^{-4}$ &                                   \\
          &                       & $\Gamma_2=2$ (f) &                            \\
          &                       & norm$_2$=$(1.76^{+0.54}_{-0.55})\times10^{-6}$ &                                   \\
\enddata
\label{extensionspectrum_s_table}
\tablecomments{(f) indicates that the parameter was frozen.}
\end{deluxetable}

\newpage

\end{document}